\documentclass[conference]{IEEEtran}
\IEEEoverridecommandlockouts
\usepackage{cite}
\usepackage{amsmath,amssymb,amsfonts,bm}
\usepackage{algorithmic}
\usepackage{graphicx}
\usepackage{textcomp}
\usepackage{xcolor}
\usepackage{caption}
\usepackage{subcaption}
\usepackage{siunitx}
\usepackage{booktabs}%
\usepackage{automated_macros}
\usepackage{todonotes}
\usepackage{nicefrac}

\def\BibTeX{{\rm B\kern-.05em{\sc i\kern-.025em b}\kern-.08em
    T\kern-.1667em\lower.7ex\hbox{E}\kern-.125emX}}

\makeatletter
\newcommand{\linebreakand}{%
    \end{@IEEEauthorhalign}
    \hfill\mbox{}\par
    \mbox{}\hfill\begin{@IEEEauthorhalign}
}
\makeatother

\graphicspath{{figures/}}

\begin{document}

\title{Towards a Hybrid Digital Twin: \\Physics-Informed Neural Networks as \\Surrogate Model of a Reinforced Concrete Beam}

\author{\IEEEauthorblockN{Tarik Sahin}
\IEEEauthorblockA{\textit{Institute for Mathematics and} \\ \textit{Computer-Based Simulation (IMCS)} \\
\textit{University of the Bundeswehr Munich}\\
Neubiberg, Germany \\
tarik.sahin@unibw.de}
\and
\IEEEauthorblockN{Daniel Wolff}
\IEEEauthorblockA{\textit{Institute for Mathematics and} \\ \textit{Computer-Based Simulation (IMCS)} \\
\textit{University of the Bundeswehr Munich}\\
Neubiberg, Germany \\
d.wolff@unibw.de}
\and
\IEEEauthorblockN{Max von Danwitz}
\IEEEauthorblockA{\textit{German Aerospace Center (DLR)} \\
\textit{Institute for the Protection of } \\
\textit{Terrestrial Infrastructures} \\ 
Sankt Augustin, Germany \\
max.vondanwitz@dlr.de}
\linebreakand 
\IEEEauthorblockN{Alexander Popp\IEEEauthorrefmark{1}\IEEEauthorrefmark{2}}
\IEEEauthorblockA{\IEEEauthorrefmark{1}\textit{Institute for Mathematics and Computer-Based Simulation (IMCS)} \\
\textit{University of the Bundeswehr Munich}\\
Neubiberg, Germany \\
\IEEEauthorrefmark{2}\textit{German Aerospace Center (DLR)} \\
\textit{Institute for the Protection of Terrestrial Infrastructures} \\
Sankt Augustin, Germany \\
alexander.popp@unibw.de}
}

\maketitle

\begin{abstract}
In this study, we investigate the potential of fast-to-evaluate surrogate modeling techniques for developing a hybrid digital twin of a steel-reinforced concrete beam, serving as a representative 
example of a civil engineering structure. As surrogates, 
two distinct models are developed utilizing physics-informed neural networks, 
which integrate experimental data with given governing laws of physics. 
The experimental data (sensor data) is obtained from a previously conducted four-point bending test. 
The first surrogate model predicts strains at fixed locations along the center line of the beam for various time instances. 
This time-dependent surrogate model is inspired by the motion of a harmonic oscillator. For this study,
we further compare the physics-based approach with a purely data-driven method, revealing the significance of physical laws 
for the extrapolation capabilities of models in scenarios with limited access to experimental data.
Furthermore, we identify the natural frequency of the system by utilizing the physics-based model as an inverse solver. 
For the second surrogate model, we then focus on a fixed instance in time and combine the sensor data with the equations 
of linear elasticity to predict the strain distribution within the beam. This example reveals 
the importance of balancing different loss components through the selection of suitable loss weights.

\end{abstract}


\section{Introduction}
A hybrid digital twin represents a digitalized version of a physical asset or a system, serving as a metamodel to combine 
experimental or simulation data with the underlying physical principles of the considered system to enhance model 
accuracy and reliability.  
Many applications of hybrid digital twins have been introduced and developed 
across a wide range of industries, including construction, manufacturing, automotive, and many others, 
see e.g. the cases in the studies by Jiang~\cite{Jiang2021} and Singh~\cite{Singh2022}.

Surrogate modeling has become a vital technique for enabling hybrid digital twins, 
providing fast-to-evaluate models that deliver real-time accurate predictions. 
Once these models are constructed and trained, they can be continuously fed with the streamed data
and can, e.g., be used to monitor the structural health of a system, which is extremely important for elements of the so-called critical infrastructure such as bridges.   
On a high level, surrogate modeling techniques can be 
classified into two main categories: purely data-driven methods and physics-informed approaches. 
Purely data-driven techniques solely rely on experimental data to identify underlying patterns in the data. 
They are also referred to as black-box models as they try to infer a system's behavior purely from input-output pairs, 
i.e., there is no information about the actual behavior of the system, 
which may result in poor predictions where data is limited. These models are mainly 
developed using machine learning and deep learning techniques, e.g., neural networks. 
On the other hand, physics-informed techniques, also known as gray-box models, 
enhance data-driven techniques by incorporating (parts of) the knowledge about the problem's governing physical 
equations into the learning process. Therefore, these methods aim to combine the strengths of 
both approaches to improve accuracy and reliability, particularly for extrapolation purposes where no data is available. 
As an example, physics-informed neural networks (PINNs)~\cite{raissi2019physics} have emerged in recent years as a highly promising approach.  

In this study, we develop fast-to-evaluate reduced-order models (ROMs) to 
predict strain distributions within a reinforced concrete beam as a representative example of civil engineering structures. 
As ROMs, two distinct models are developed based on PINNs. 
Starting with a time series of measurements from a four-point bending test, we first construct a ROM to examine the temporal evolution of strains at a fixed position inside the beam.
First, we leverage a purely data-driven approach by training a neural network solely with the available experimental data. 
However, since the beam is subject to a periodic loading, we then exploit knowledge about the motion of a harmonic oscillator and utilize a PINN to learn the system's behavior.
This comparison shows the benefits of including physical principles with respect to the model's prediction capability when no experimental data is available. 

The second ROM then focuses on capturing spatial effects for a fixed instance in time. Therefore, 
we integrate the collected experimental data with the equations of linear elasticity to predict strains inside the beam. We demonstrate that 
including experimental data from both the compression and tension sides of the beam improves the prediction accuracy.  
However, this inclusion may result in non-smooth and sensitive model results due to measurement noise in the experimental data.
To tackle this problem, we tune the loss weights of the deployed PINN to find a good balance between the physics-based loss and the experimental data loss. 

The remainder of this study is structured as follows: Section~\ref{sec:HDT} introduces the concept of hybrid digital 
twins including a proof of concept and the investigated physical asset, i.e., a reinforced concrete beam. 
In Section~\ref{sec:methods}, we recapitulate the general concept of PINNs as forward solvers for (partial) 
differential equations. Sections~\ref{sec:temporalROM}~and~\ref{sec:spatialROM} introduce the developed ROMs to 
capture the temporal and spatial behavior of the system including corresponding physical laws, 
tailored PINN formulations, and results including distinct scenarios.
Section~\ref{sec:conclusion} concludes the article by summarizing our main findings and 
proposing future research directions.

\section{The concept of hybrid digital twins}\label{sec:HDT}
A digital twin is an up-to-date virtual representation of a physical asset, system, or process to precisely simulate 
the characteristics and behaviors of corresponding real-world objects \cite{grieves2014digital}. In other words, it can be considered as a mapping 
between the physical realm and the virtual realm. On the other hand, a hybrid digital twin\cite{chinesta2020} merges physics-based modeling 
(virtual twin) with data-driven methods (digital twin) to create simulation tools capable of monitoring, identifying underlying 
trends, incorporating real-time operational data, making predictions, and even further conducting what-if simulations. Specifically, in the context 
of bridges, digital twins can play a vital role in condition and lifetime assessment, facilitating structural health monitoring (SHM), 
structural maintenance, and increasing the system's overall resilience \cite{torzoni2024}.  

\begin{figure}[htbp]
    \includegraphics[width=0.475\textwidth]{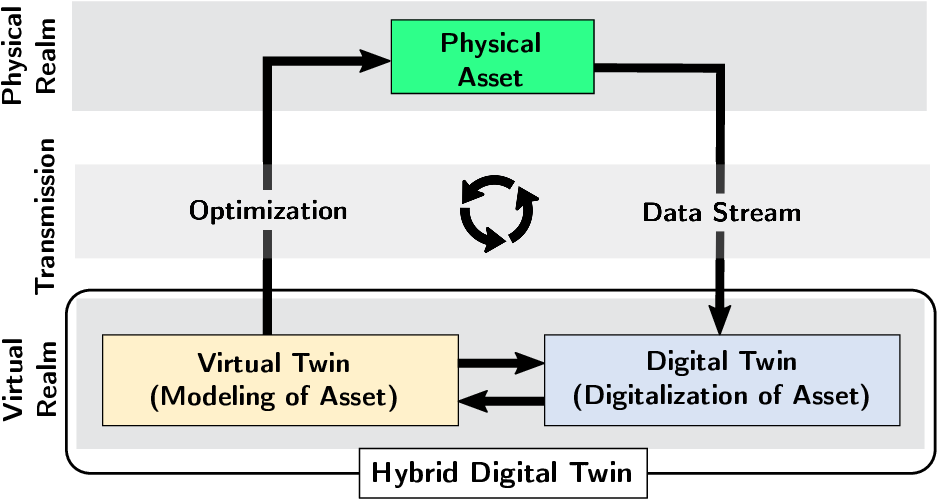}
    \caption{The conceptual workflow between hybrid digital twin, virtual twin, and digital twin.}
    \label{figHDTDT}
\end{figure}

The conceptual workflow between hybrid digital twin, virtual twin, and digital twin is depicted in Fig.~\ref{figHDTDT}. The physical
asset lies in the physical realm. The data generated by the physical asset is streamed to the digital twin.
Fast, efficient, and secure data transmission methods between the physical and virtual realms, such as file transfer and streaming protocols, 
are of paramount importance. 
The digital twin represents the digitalization of the system based on data-driven methods, such as machine learning and deep learning techniques. 
On the other hand, the virtual twin is constructed through physics-based modeling of the system.
Both the digital twin and the virtual twin exist within the virtual realm and together constitute the hybrid digital twin. 
\subsection{Proof of Concept}
In this contribution, we aim to generate a hybrid digital twin concept for steel-reinforced concrete beams 
as a representative component of bridges or, more generally, civil engineering structures. 

\begin{figure}[htbp]
    \includegraphics[width=0.475\textwidth]{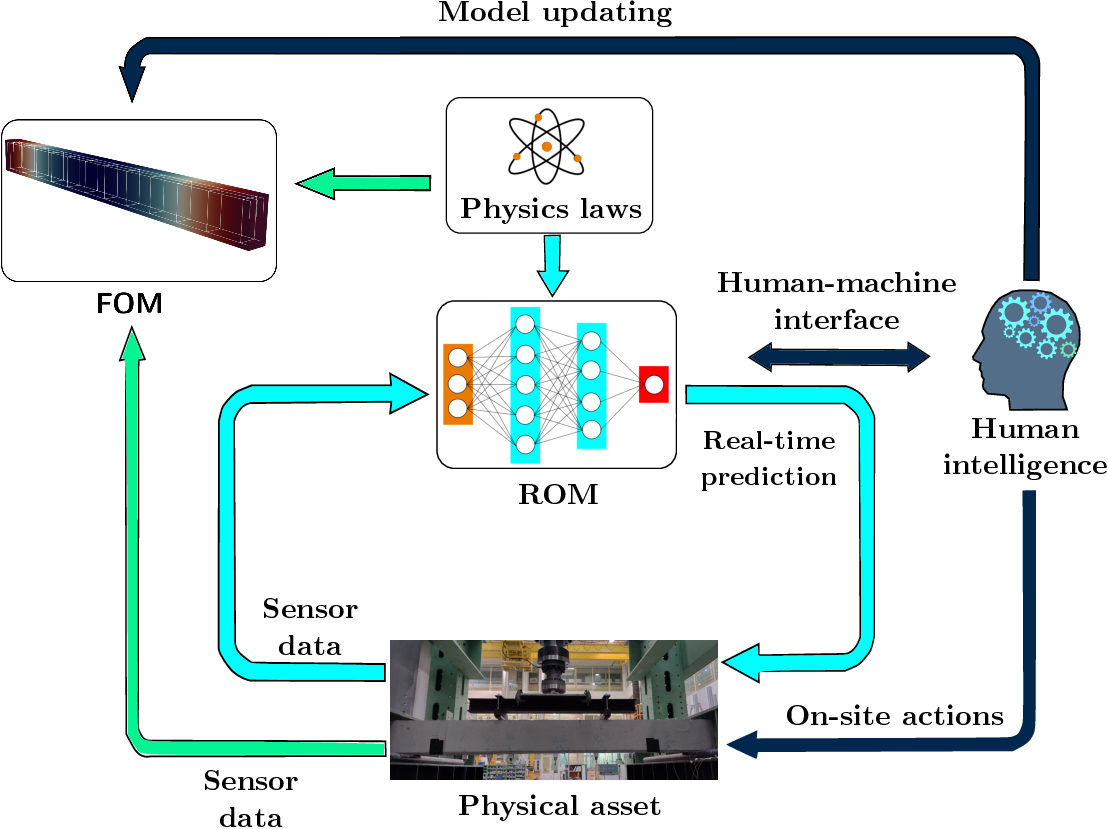}
    \caption{A hybrid digital twin concept for the steel-reinforced concrete beam.}
    \label{figHDTConcrete}
\end{figure}

As shown in Fig.~\ref{figHDTConcrete}, a steel-reinforced concrete
beam is defined as the physical asset. Following a four-point bending test, the collected sensor data is transmitted to 
a physics-based full-order model (FOM) as well as a data-driven reduced-order model (ROM) which is fast-to-evaluate 
and enhanced with physics laws, also known as so-called gray-box modeling. 
The fast evaluation capability of the data-driven ROM allows real-time predictions. 
Even with the inclusion of physics, gray-box modeling techniques require human intelligence to manage the entire framework, 
as the reliability and accuracy of data-driven methods remain topics of debate. 
Therefore, considering the interaction among models, human intelligence ultimately determines the actions on-site to be taken. 
As for the ROM, we develop fast-to-evaluate models utilizing physics-informed neural networks. 
The FOM is not within the scope of this study and for further details we refer to \cite{von2023hybrid}.
\subsection{Physical Asset: Reinforced Concrete Beam}
\begin{figure}[htbp]
    \center
    \includegraphics[width=0.485\textwidth]{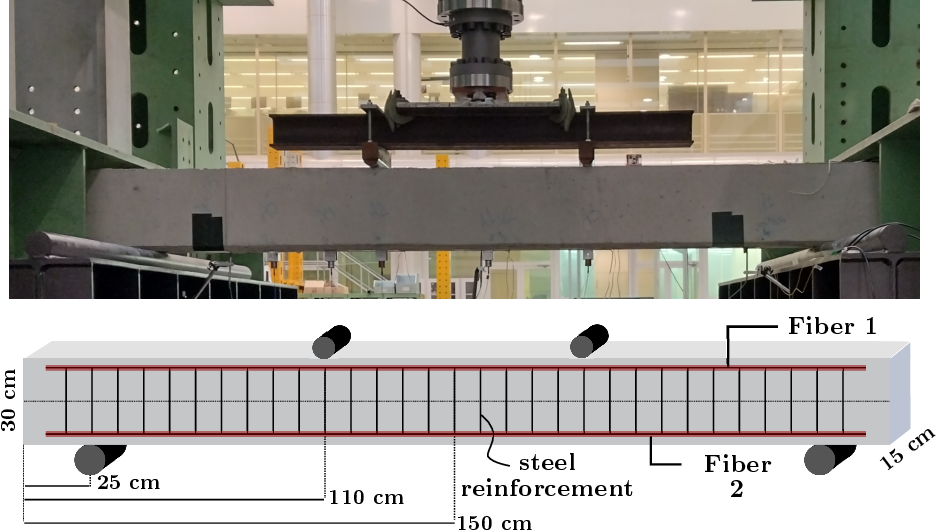}
    \caption{Four-point bending test of the reinforced concrete beam including a schematic drawing. Photo taken from \cite{von2023hybrid}.}
    \label{figBeam}
\end{figure}

In this study, a steel-reinforced concrete beam as a representative component in civil engineering structures
is defined as the physical asset of the conceptual workflow.
The dimensions of the constructed beam are given in Fig.~\ref{figBeam}.
The concrete has a Young's modulus $E_c = \SI{29}{\GPa}$, while the construction steel has a 
Young's modulus $E_s = \SI{200}{\GPa}$. For the detailed specifications of the considered beam, 
we refer to \cite{braml2022b,von2023hybrid}.

\begin{figure}[htbp]
    \includegraphics[width=0.475\textwidth]{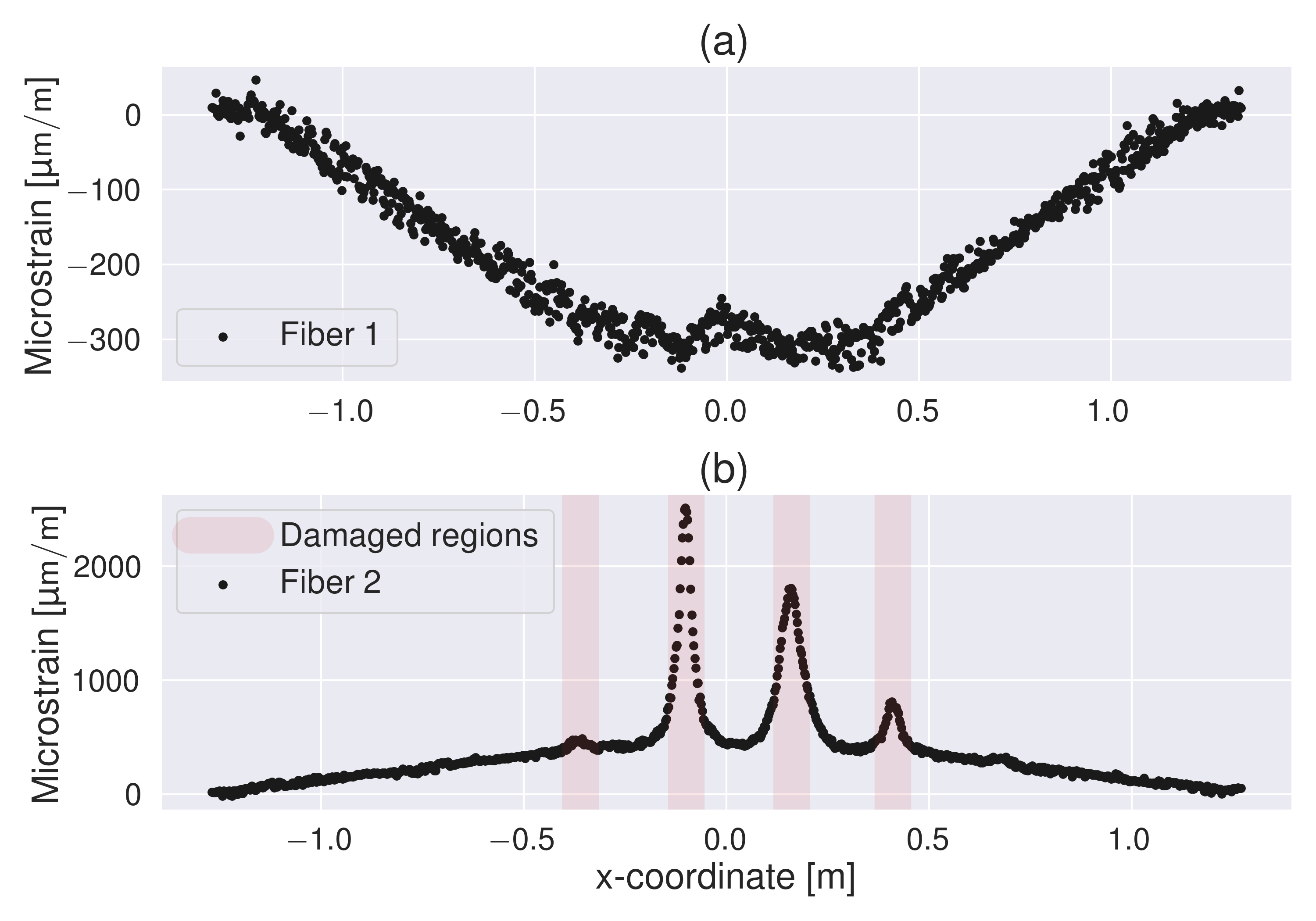}
    \caption{Strain measurements from fiber-optical sensors are captured under a peak load of $\SI{10}{kN}$. 
    (a) Fiber 1 on the compression side, 
    (b) fiber 2 on the tension side including damaged regions.}
    \label{figMeasurements}
\end{figure}

A four-point bending test is conducted to generate the dataset for our workflow. Fiber-optical sensors are 
embedded into the concrete beam to measure longitudinal strains (microstrains) along the beam axis, 
with two sensors positioned on the compression side and four sensors located on the tension side. 
However, for this study, only two channels are considered: one from the compression side (fiber 1) 
and one from the tension side (fiber 2). 
For the loading, a sinusoidal cyclic load with a peak of $\SI{10}{kN}$ is applied. 
Each fiber-optical sensor records roughly 1000 data points as demonstrated in Fig.~\ref{figMeasurements}.
While the compression side exhibits the anticipated bending behavior, 
cracks are identified on the tension side, which weaken the beam locally. 
This structural weakening of the material results in a drastic increase of the 
microstrains (Fig.~\ref{figMeasurements}b).

\begin{figure}[htbp]
    \center
    \includegraphics[width=0.485\textwidth]{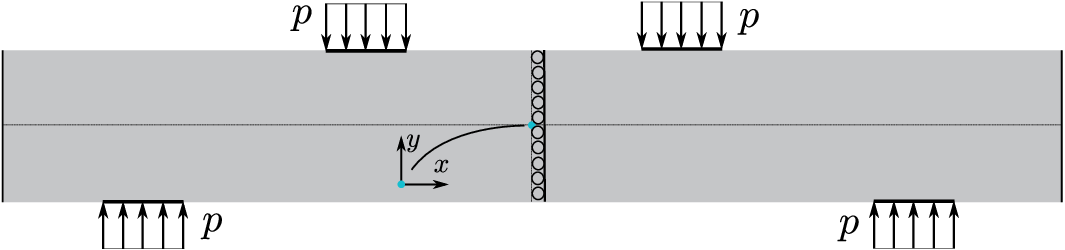}
    \caption{A relaxed model representation of the four-point bending test.}
    \label{figSimplifiedBeam}
\end{figure}

Furthermore, a relaxed model of the experimental test is depicted in Fig.~\ref{figSimplifiedBeam}. 
Here, supports are replaced by traction boundary conditions, and the beam is fixed in x-direction at its center.
In the following sections, this relaxed model serves as the foundation for the second ROM.

\section{Reduced Order Modeling: Physics-Informed Neural Networks}\label{sec:methods}

The main idea of PINNs is to approximate the solution 
of a given (partial) differential equation (system) with a neural network by incorporating the governing 
physical constraints into the loss function.

A general formulation for partial differential equations (PDEs) can be presented in residual form along with 
corresponding initial and boundary conditions (BCs) as
\begin{equation}
    \begin{aligned}
    \label{eq:governing}
    \mathcal{R}\bigl[ \latentsol (\pdeInput, \timeInput) \bigr]&=\boldsymbol{0} \quad \text{on} \ \Omega, \\
    \mathcal{B}\bigl[ \latentsol (\pdeInput, \timeInput) \bigr]-\boldsymbol{g}(\pdeInput, \timeInput)&=\boldsymbol{0} \quad \text{on} \ \partial \Omega,
    \end{aligned}
\end{equation}
where $\mathcal{R}[\cdot]$ represents a differential operator, $\boldsymbol{u}$ is the unknown solution, 
spatial coordinates denoted by $\pdeInput$ and temporal coordinates denoted by $t$ span the space-time domain $\Omega$ with boundary $\partial \Omega$, 
$\mathcal{B}[\cdot]$ is the boundary and initial condition operator, and  
$\boldsymbol{g}$ represents defined boundary and initial conditions. To approximate $\boldsymbol{u}$, a fully-connected neural 
network with $L$ layers is defined as \cite{lawrence1993} 
\begin{equation}
    \label{eq:approx}
    \latentsol \approx \pinnsol := (\netOutput)^{'}, \quad  \netOutput : \mathbb{R}^{d} \rightarrow \mathbb{R}^{n}.
\end{equation}
Here, the output of the neural network is denoted as $\netOutput$, where
$\boldsymbol{\theta}$ represents a set of trainable parameters, namely the network's weights and biases.
The function $(\cdot)'$ signifies a customized output transformation, elaborated extensively in the work of 
Sahin et al.~\cite{sahin2024}. The input dimension is denoted as $d$ and the 
output dimension is represented by $n$.
In our case, the network input $\netInput$ consists of the space-time coordinates, 
i.e. $\netInput = (\pdeInput, \timeInput)$. However, the input of the network is not limited to 
space-time coordinates and could additionally comprise (variable) model parameters.  
The output of every layer is passed through a nonlinear activation function, denoted as $\psi$, before being processed by the next layer.

To ensure that $\pinnsol$ is a reliable approximation of $\latentsol$, the network parameters must be optimized to fulfill the given constraints, i.e.,
\begin{equation}
    \boldsymbol{\theta}^*=\underset{\boldsymbol{\theta}}{\arg \min } \ \mathcal{L}(\boldsymbol{\theta}),
\end{equation}
where 
\begin{equation}
    \mathcal{L}(\boldsymbol{\theta}) = w_{1} \mathcal{L}_{\mathrm{PDEs}}+
    w_{2} \mathcal{L}_{\mathrm{BCs}}+
    w_{3} \mathcal{L}_{\mathrm{ICs}} +
    w_{4} \mathcal{L}_{\mathrm{EXPs}}.
\end{equation}
Here, the overall loss $\mathcal{L}$ is composed of PDE losses $\mathcal{L}_{\mathrm{PDEs}}$, boundary condition losses 
$\mathcal{L}_{\mathrm{BCs}}$, initial condition losses $\mathcal{L}_{\mathrm{ICs}}$, and experimental data losses $\mathcal{L}_{\mathrm{EXPs}}$. 
The $w_i$ denote individual loss weights assigned to each loss component, which improves the convergence of the optimization process.

In the following sections, we propose two ROMs based on PINNs.  
The main difference between these models lies in the underlying physical principles.
The first ROM focuses on the temporal strain response of the system
enhanced by the motion of a simple harmonic oscillator. 
In the second approach, the ROM predicts the displacements and stresses within the spatial domain of the concrete beam, which is modeled with 
the equations of linear elasticity. The introduced PINN models are referred to as temporal ROM and spatial ROM,
respectively.

\section{A ROM for Capturing the Temporal Response}
\label{sec:temporalROM}

The first aim of this paper is to develop a model that can accurately capture the temporal behavior of the system under consideration. Therefore, we utilize measurement data that was collected at a fixed location, specifically the center of fiber~1 (cf. Fig.~\ref{figBeam}) across various time instances.
\begin{figure}[htbp]
    \center
    \includegraphics[width=0.48\textwidth]{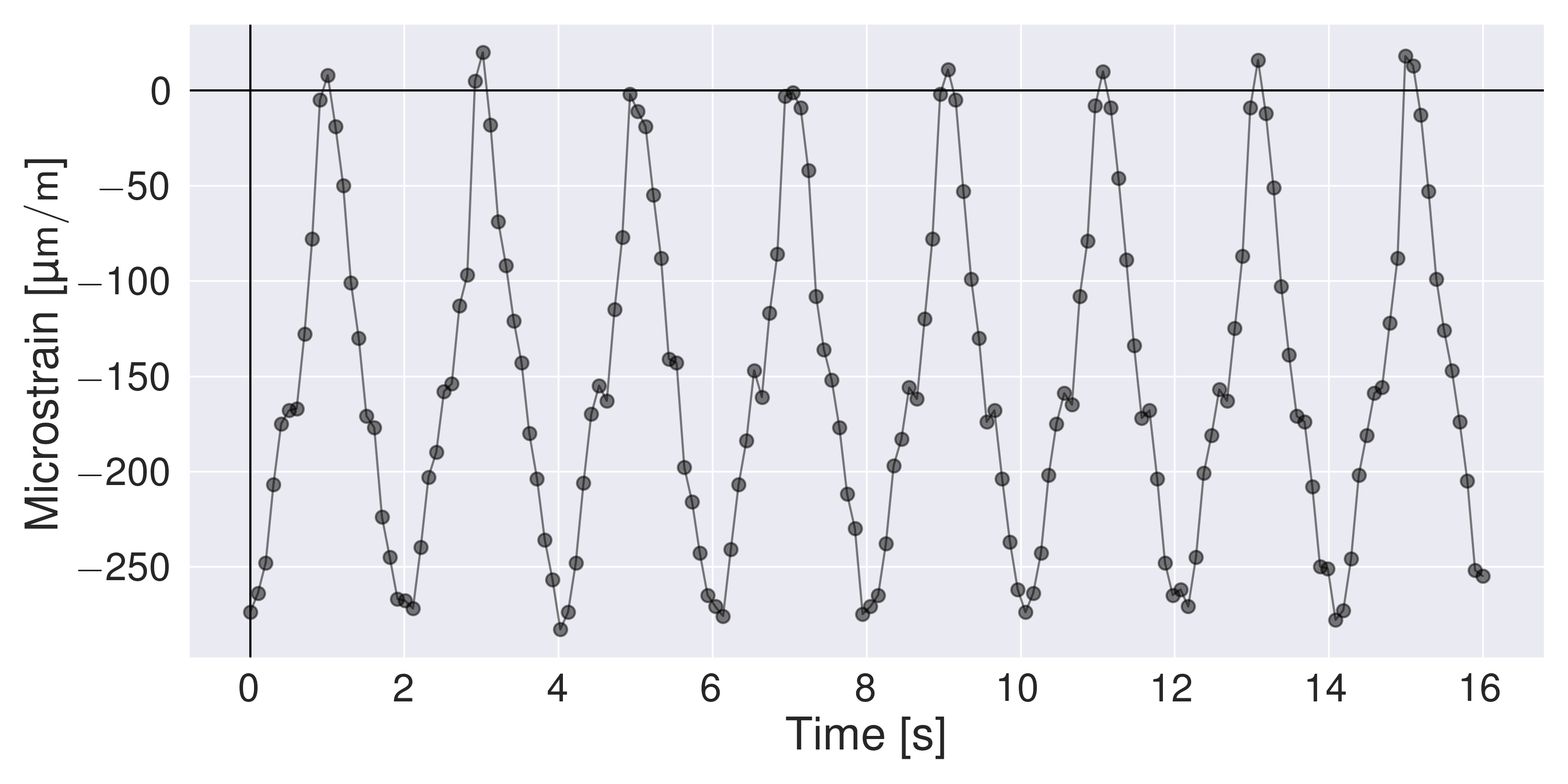}
    \caption{Strain measurements recorded from the center of fiber~1 over time.}
    \label{fig:TimeData}
\end{figure}

Fig.~\ref{fig:TimeData} shows these measured strains for the first $\SI{16}{\second}$. Each measurement interval of one second contains ten individual measurement points.
Due to the sinusoidal loading, the measured strain data points also follow a harmonic behavior in this example. In general, however, the underlying data generation process is unknown. 
To prepare for such a more realistic scenario, the next section introduces the ordinary differential equation (ODE) for a harmonic oscillator, which can be used to describe a data distribution as shown in Fig.~\ref{fig:TimeData}. While for this simple introductory example we are still able to solve the differential equation analytically, this is no longer possible for more complicated cases, as we will show in Section~\ref{sec:spatialROM}.
\subsection{Motion of a Simple Harmonic Oscillator}
The balance of forces, governed by Newton's second law, for a simple harmonic oscillator is expressed as
\begin{equation}
    \label{eq:harmonic}
    m \frac{d^2 u}{dt^2} = -k u \quad \Leftrightarrow \quad  \frac{d^2 u}{dt^2} + \omega_0^2 u = 0.
\end{equation}
Here, $u$ represents the displacement from the equilibrium position $x_0$, $k$ is the spring constant, 
$m$ is the inertial mass, and $\omega_0=\sqrt{\nicefrac{k}{m}}=2 \pi f$ is the natural oscillating frequency.
The longitudinal strain $\varepsilon$ can be determined from the displacement and the equilibrium position as
$\varepsilon=\nicefrac{u}{x_0}$.

\subsection{A PINN-based temporal ROM}
\begin{figure}[htbp]
    \center
    \includegraphics[width=0.48\textwidth]{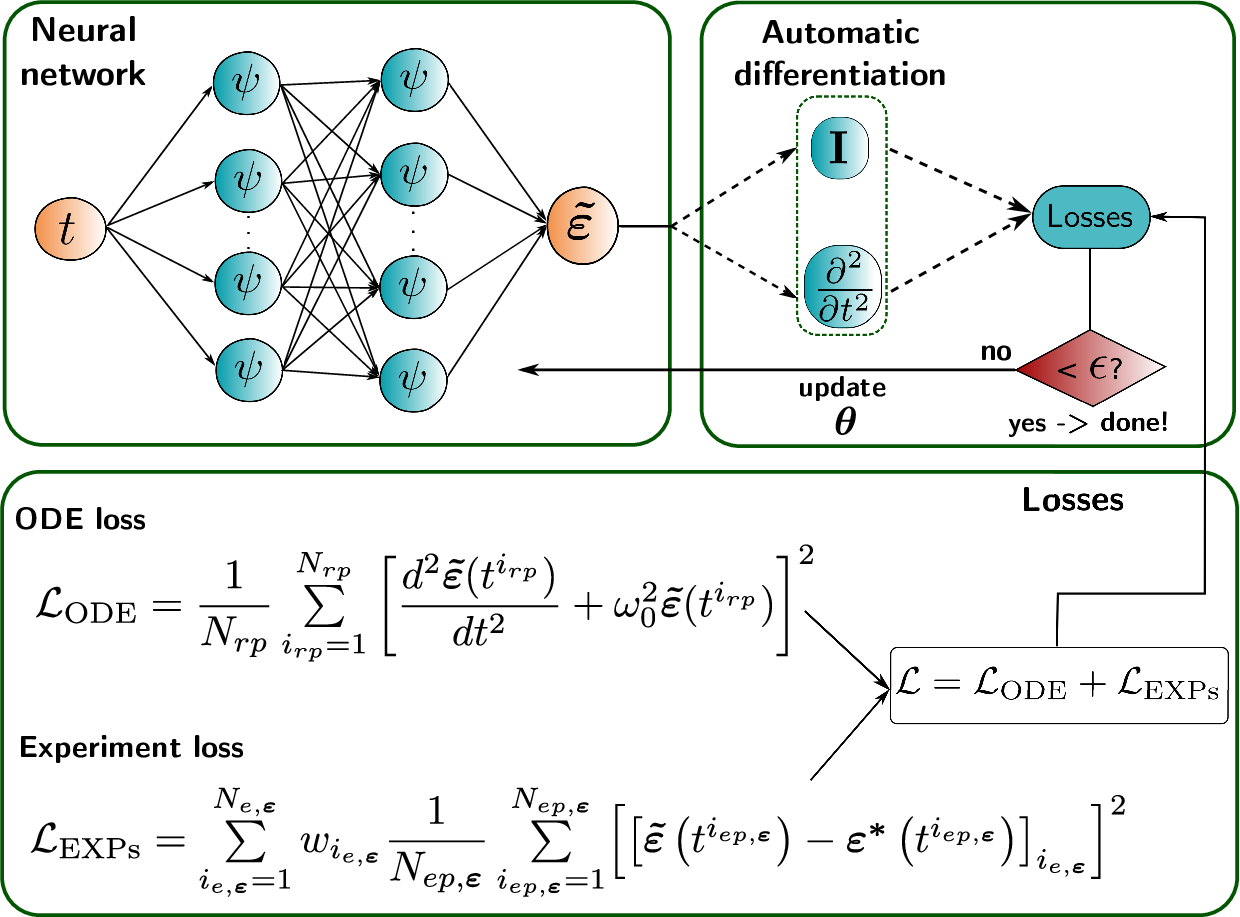}
    \caption{Detailed schematic of the temporal ROM. }
    \label{fig:ROMtime}
\end{figure}

Inspired by a harmonic oscillator, a fully-connected neural network (FNN) approximates the 
longitudinal strain within a selected fiber as
\begin{equation}
    \tilde{\varepsilon} := \mathcal{N}_{\varepsilon}(t;\boldsymbol{\theta}).
\end{equation}
The time $t$ is the neural network input, and the strain in the longitudinal direction of the beam, 
$\tilde{\varepsilon}$, serves as the network output. 
With the help of automatic differentiation (AD), second-order derivatives are calculated. 
The total loss includes $\mathcal{L}_{\mathrm{ODE}}$ and $\mathcal{L}_{\mathrm{EXPs}}$. 
Here, $\mathcal{L}_{\mathrm{ODE}}$ represents the loss due to the ODE defined in Eq.~\eqref{eq:harmonic}.
As the measurement data already includes an IC, i.e., $\varepsilon(t=0)=-\SI{293}{\micro\meter}$, we exclude the loss component $\mathcal{L}_{\mathrm{ICs}}$.
Although we can determine an approximate value for the square of the oscillating frequency of $\omega_0^2\approx \SI{9.87}{\radian\squared\Hz\squared}$ from the measurement data, we consider the parameter to be unknown in our experiments and let the PINN infer it from the provided data. The detailed workflow is shown in Fig.~\ref{fig:ROMtime}. For the network, we employ a fully-connected feed-forward neural network with 3 layers consisting of 30 neurons each. The $\tanh$ activation function is chosen.

Regarding the training, we first use the stochastic gradient descent optimizer \textit{Adam} with a learning rate of 
0.001 for 5000 epochs. After that, we switch to the limited memory BFGS algorithm, 
which incorporates box constraints (\textit{L-BFGS-B}), and further refine the pre-trained model.  
Our workﬂow is built upon the DeepXDE package~\cite{lu2021deepxde} and we refer to the
DeepXDE documentation for the default settings of the \textit{L-BFGS-B} optimizer.

\subsection{Results}
\label{sec:usecases}

In the following, we consider two scenarios. In the first scenario, a purely data-driven neural network is trained on the 
time-dependent experimental data from the compression side of the beam (Fig.~\ref{fig:TimeData}) 
without incorporating any physics. 
The specifications for this network remain the same as those of the temporal ROM, except that $\mathrm{sin}$ is used as activation function. 
In the second scenario, we employ the temporal ROM instead, enhancing the learning process with the physical knowledge provided by Eq. \eqref{eq:harmonic}.

\begin{figure}[htbp]
    \center
    \includegraphics[width=0.45\textwidth]{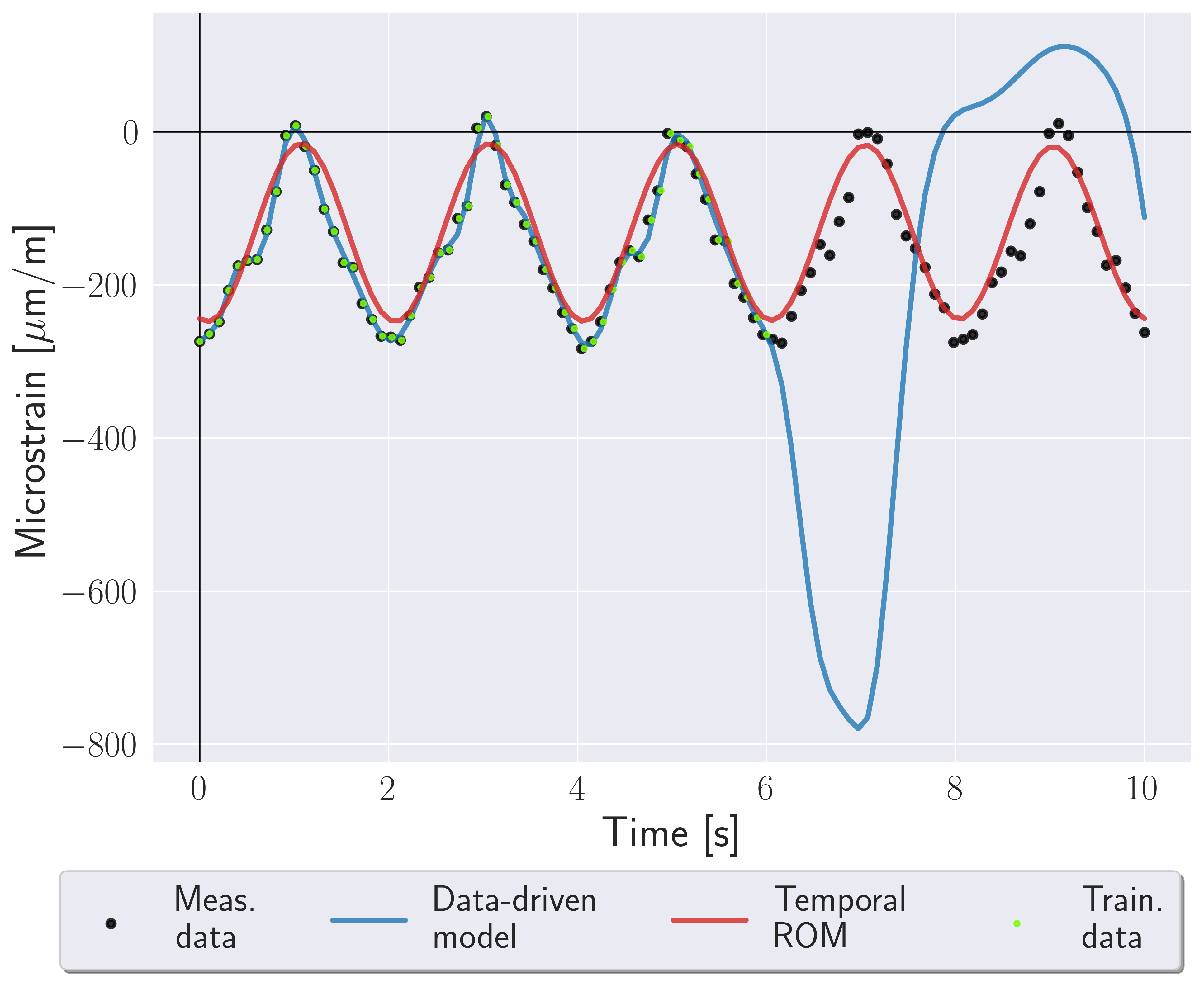}
    \caption{Comparison of a purely data-driven model and the temporal ROM
    including training and measurement data points. }
    \label{fig:ROM_2_Results}
\end{figure}

Fig.~\ref{fig:ROM_2_Results} compares the results of the purely data-driven neural network and the temporal ROM.
Both models are trained using the first $\SI{6}{\second}$ of the measurement data as training data. 
While the data-driven model perfectly aligns with the training data,
the predictions of the temporal ROM do not align perfectly. 
This discrepancy stems from the fact that the sensor data contains measurement noise, which is not accounted for in the physical laws.
However, the main advantage of the temporal ROM becomes apparent after $t=\SI{6}{\second}$:
While the data-driven model diverges extensively from the measurement 
data, the temporal ROM still captures the trend of the measurements correctly.
This shows that the purely data-driven model does not generalize well to cases, where no measurement data is available.
Due to the inclusion of physical laws, the temporal ROM, however, is still able to make decent predictions even in a no-data regime.

With respect to the training times, the purely data-driven model trains for $\SI{10.2}{\second}$ with the \textit{Adam} optimizer,  
and additional $\SI{12.7}{\second}$ with the \textit{L-BFGS-B} optimizer. The training of the temporal ROM
takes approximately $\SI{12.6}{\second}$ for the iterations with \textit{Adam}, and $\SI{23}{\second}$ 
for \textit{L-BFGS-B}. The prediction times are quite similar, taking only $\SI{0.003}{\second}$ and $\SI{0.004}{\second}$ 
for the purely data-driven model and the temporal ROM, respectively. 

\begin{figure}[htbp]
    \center
    \includegraphics[width=0.45\textwidth]{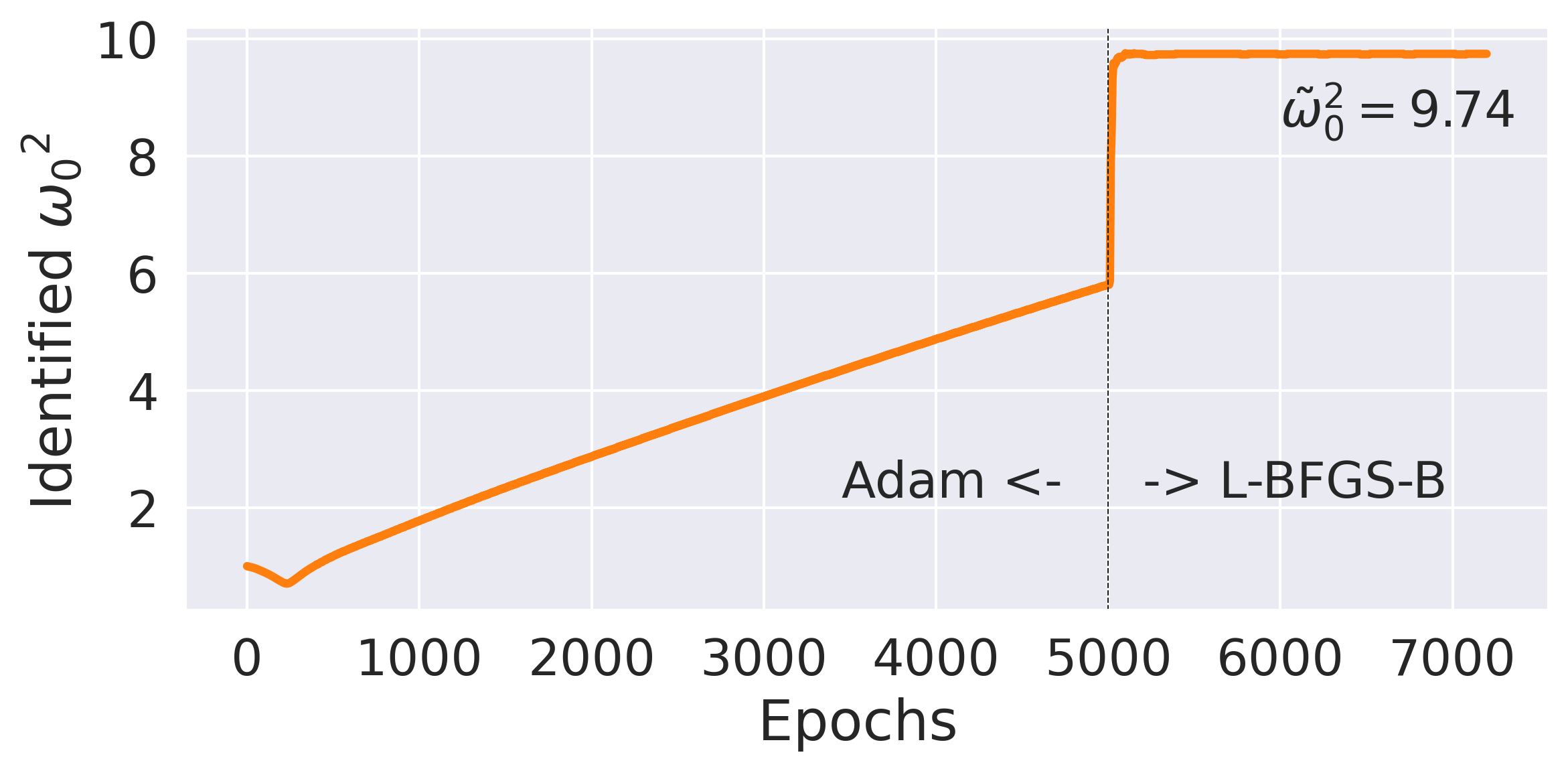}
    \caption{Identification of $\omega_0$.}
    \label{fig:inverse}
\end{figure}

As mentioned in the previous section, for the case of the temporal ROM, we identify $\omega_0$ by 
additionally exploiting the PINN's capability as an inverse solver. Therefore, we add $\omega_0$ to the set of trainable
network parameters $\boldsymbol{\theta}$ and start from an initial guess of 1. As shown in Fig.~\ref{fig:inverse},
the natural frequency of the system is identified as $\tilde{\omega}_0^2\approx\SI{9.74}{\radian\squared\Hz\squared}$,
which is sufficiently close to the value of $\omega_0^2 \approx \SI{9.87}{\radian\squared\Hz\squared}$ computed from the measurements.
Moreover, it can be seen that applying \textit{L-BFGS-B} after \textit{Adam} drastically improves the convergence. 

\section{A ROM for Capturing the Spatial Behavior}
\label{sec:spatialROM}

After creating a ROM for the temporal response in the previous section, this section now introduces PINN-based ROMs that can predict the spatial strain distribution inside the beam. For simplicity, we model the beam in the following as a two-dimensional rectangular domain and neglect its depth. Note that finding a physical model which accurately describes the physical behavior of the steel-reinforced concrete beam is a research topic of its own. In the following, we therefore employ a strongly simplified physical model and show that this produces sufficiently accurate results in combination with the available measurement data. 

\begin{figure*}[htbp]
    \center
    \includegraphics[width=0.8\textwidth]{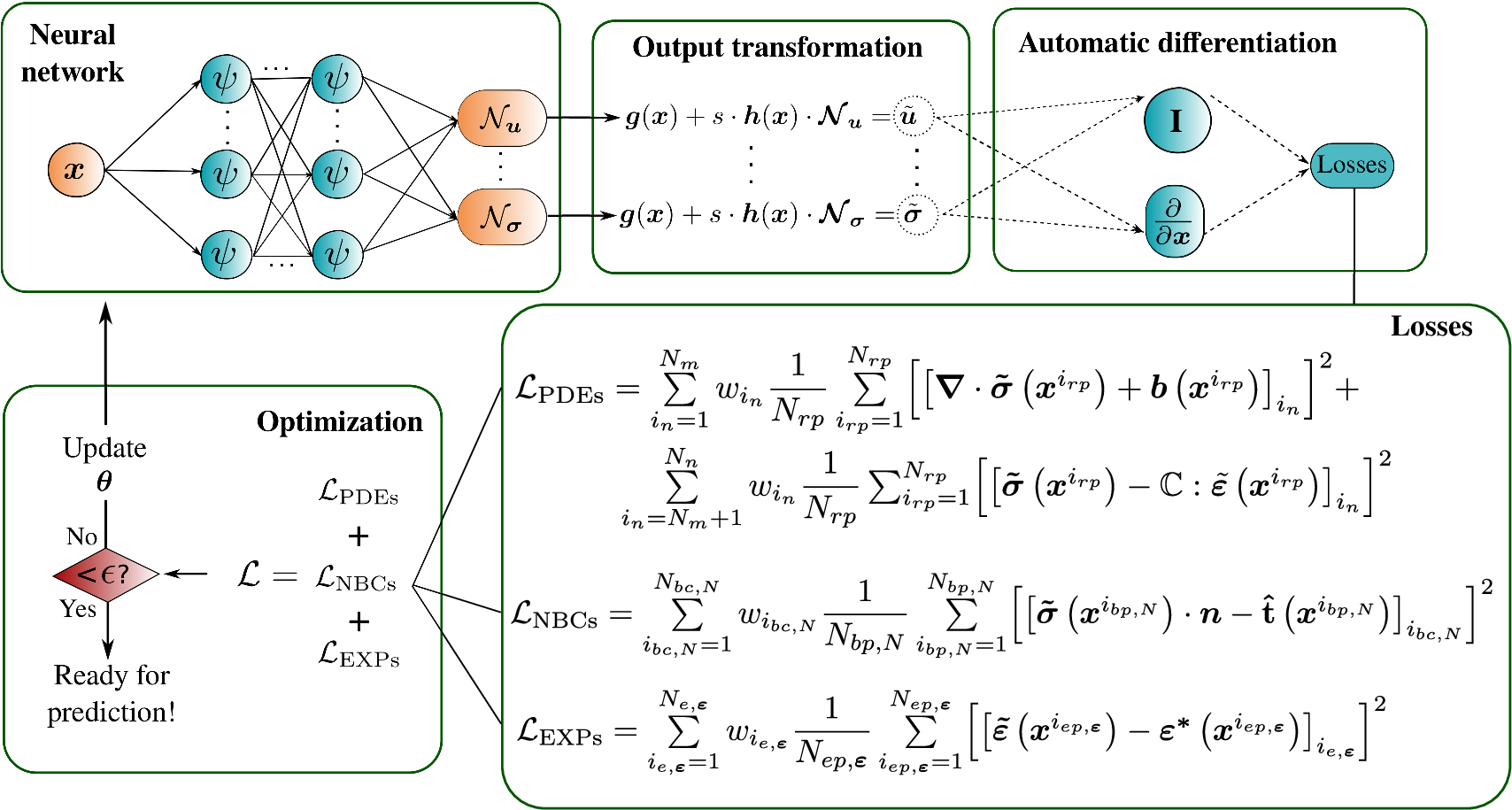}
    \caption{Detailed schematic of the spatial ROM. 
    A PINN learns to solve the equations of linear elasticity using a mixed-variable formulation.
    Additional loss components stem from the Neumann BCs and the experimental data loss.}
    \label{fig:ROMspace}
\end{figure*}
\subsection{Equations of Linear Elasticity}
The boundary value problem for the Cauchy momentum balance equation~\eqref{eq:lin_equi} is given as
\begin{align}
    \label{eq:lin_equi}
    \DivergenceOperator{\Stress} + \bodyforce & = \ZeroVector \hspace{0.9cm} \text{in } \InDomain, \tag{BE} \\ 
    \Displacement &=\boldsymbol{\hat u}  \hspace{0.85cm} \text{on } \DirichletBoundary, \tag{DBC} \\ 
    \label{eq:nbc}
    \Stress \cdot \BoundaryNormal &= \Traction \hspace{0.95cm} \text{on } \NeumannBoundary, \tag{NBC}
\end{align}
where $\Stress$ represents the Cauchy stress tensor, $\Displacement$ stands for the displacement vector, 
$\bodyforce$ denotes the body force vector, and $\boldsymbol{n}$ is the unit outward normal vector.
Prescribed displacements are denoted by $\boldsymbol{\hat u}$ on $\DirichletBoundary$ 
and $\Traction$ denotes prescribed tractions on $\NeumannBoundary$. 
The kinematic equation~\eqref{eq:compatibility} and constitutive equation~\eqref{eq:hooke}
for the deformable body are expressed as:
\begin{align}
    \label{eq:compatibility}
    \Strain &= \nicefrac{1}{2}(\NablaOperator \Displacement + \NablaOperator \Displacement^T), \tag{KE} \\
    \label{eq:hooke}
    \Stress &= \Compliance : \Strain. \tag{CE}
\end{align}
Here, $\Strain$ denotes the infinitesimal strain tensor, while $\Compliance$ represents the fourth-order elasticity tensor.
In this work, we use a linear elastic isotropic material, for which the constitutive equation can be described using Hooke's law as
\begin{equation}
    \Stress = \lambda \tr(\Strain)\mathbf{I}+2\mu\Strain,
\end{equation}
where $\lambda$ and $\mu$ are the Lam\'e parameters, $\tr(\cdot)$ denotes the trace operator, and 
$\mathbf{I}$ is the identity tensor. 

\subsection{A PINN-based spatial ROM}
The spatial ROM is formulated based on the equations of linear elasticity using a mixed-variable formulation, i.e.,
an FNN maps the spatial coordinates $\boldsymbol{x}$ onto both, the displacement vector $\boldsymbol{u}$ and 
stress tensor $\boldsymbol{\sigma}$ such that
\begin{equation}
    \label{eq:approx_mixed}
    \pinnUVector := (\mathcal{N}_{\boldsymbol{u}}(\boldsymbol{x};\boldsymbol{\theta}))^{'} 
    \hspace*{0.3cm} \text{and} \hspace*{0.3cm}
    \pinnSVector := (\mathcal{N}_{\boldsymbol{\sigma}}(\boldsymbol{x};\boldsymbol{\theta}))^{'}. \\
\end{equation}

As shown in Fig.~\ref{fig:ROMspace}, the $\mathcal{L}_{\mathrm{PDEs}}$ term is constructed in a composite
form to fulﬁll Eqs. \eqref{eq:lin_equi}, \eqref{eq:compatibility}, and \eqref{eq:hooke}. 
Note that only first-order derivatives of neural network outputs are necessary due to the mixed-variable approach. 
Spatial derivatives are again obtained using AD.

\begin{figure}[htbp]
    \center
    \includegraphics[width=0.48\textwidth]{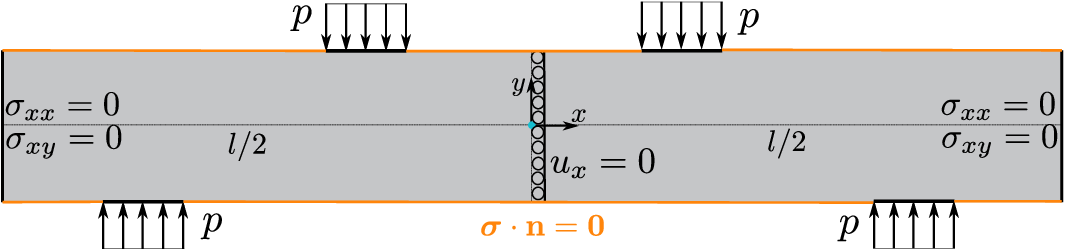}
    \caption{Boundary conditions of the relaxed system (Fig.~\ref{figSimplifiedBeam}).}
    \label{fig:BCs}
\end{figure}

Fig.~\ref{fig:BCs} visualizes the BCs of the relaxed model representation of the system introduced in Fig.~\ref{figSimplifiedBeam}: 
The beam is fixed at $x=0$ in x-direction, i.e. $u_{x}=0$. 
Since both displacement and stress components are directly deﬁned as network outputs, some BCs can be imposed as hard constraints using an output transformation. 
Based on the relaxed model representation, the following 
output transformations are applied 
\begin{equation}
    \begin{aligned}
        \pinnUComponent{x} = x & \mathcal{N}_{u_x}, \quad
        \pinnSComponent{xx} = \left(\nicefrac{l}{2}-x\right)\left(\nicefrac{l}{2}+x\right) \mathcal{N}_{\sigma_{xx}}, \\
        &\pinnSComponent{xy} = \left(\nicefrac{l}{2}-x\right)\left(\nicefrac{l}{2}+x\right) \mathcal{N}_{\sigma_{xx}},
    \end{aligned}
\end{equation}
i.e., the constraint on the displacement of the beam's center and the traction boundary conditions on the left and right edges are strongly enforced.
For details on how to formulate the output transformation, we refer to Sahin et. al~\cite{sahin2024}. 
The traction boundary conditions on the upper and lower edges are enforced as soft constraints via additional loss terms. 

Since the experimental dataset consists of longitudinal strain measurements, the experimental data loss 
$\mathcal{L}_{\mathrm{EXPs}}$ can contain both compression and tension data. 
For the test cases, 200 data points from the compression side (fiber 1) and 100 data points from the 
{tension side (fiber 2) are used. 

For the spatial ROM, a fully-connected neural network consisting of 
4 hidden layers with 50 neurons each is employed and $\mathrm{tanh}$ is chosen as the 
activation function. Since the problem is modeled in 2D, $x$ and $y$ are the network inputs, 
while $\pinnUComponent{x}, \pinnUComponent{y}, \pinnSComponent{xx}, 
\pinnSComponent{yy}, \pinnSComponent{xy}$ form the network outputs.

Similar to the temporal ROM, we initially train our neural network using the \textit{Adam} optimizer for 2000 epochs 
before switching to \textit{L-BFGS-B} to further refine the pre-trained model.

\begin{figure*}[htbp]
    \centering
    \includegraphics[width=.34\textwidth]{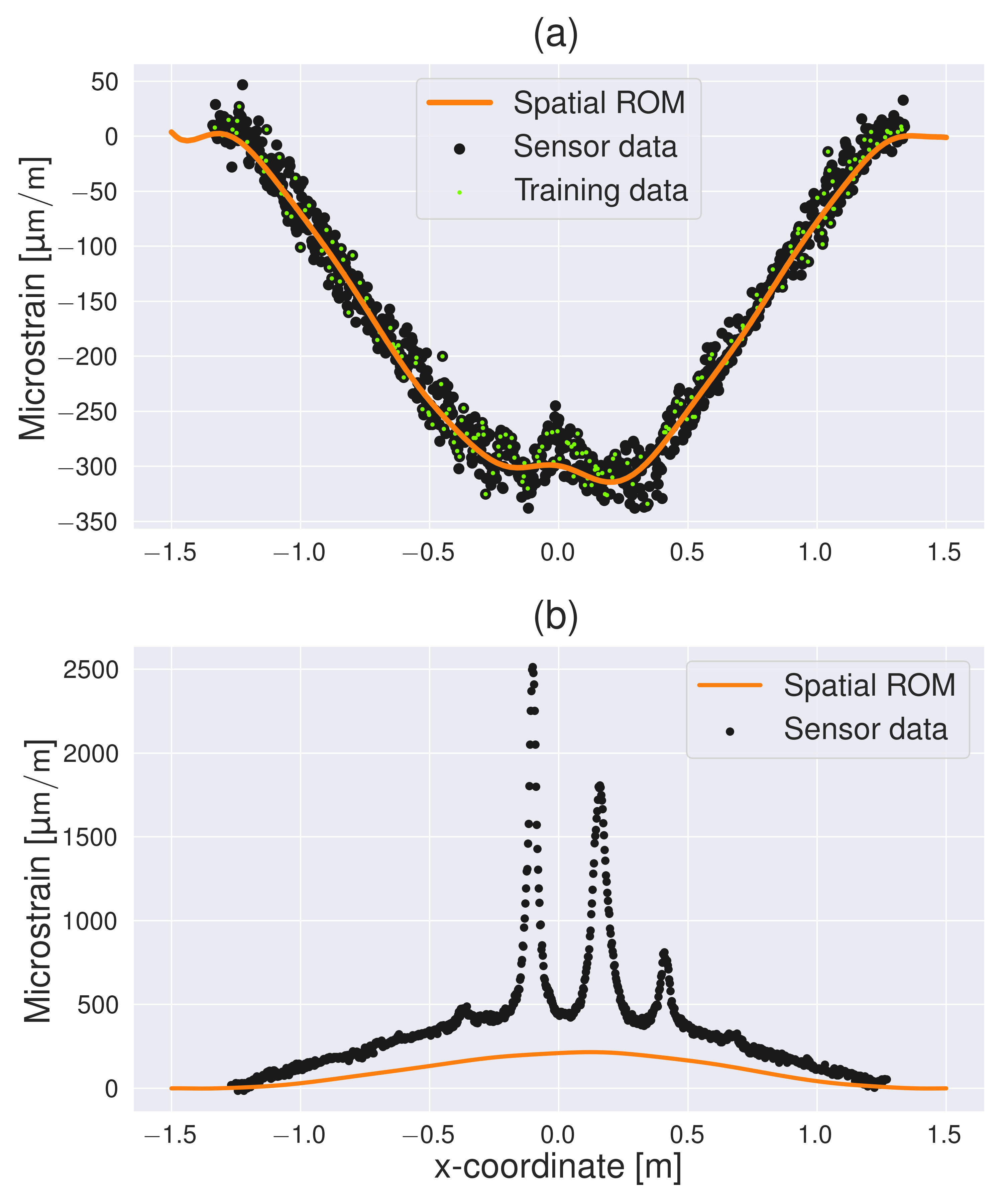}
    \includegraphics[trim={1.5cm 0 0 0},clip, width=.32\textwidth]{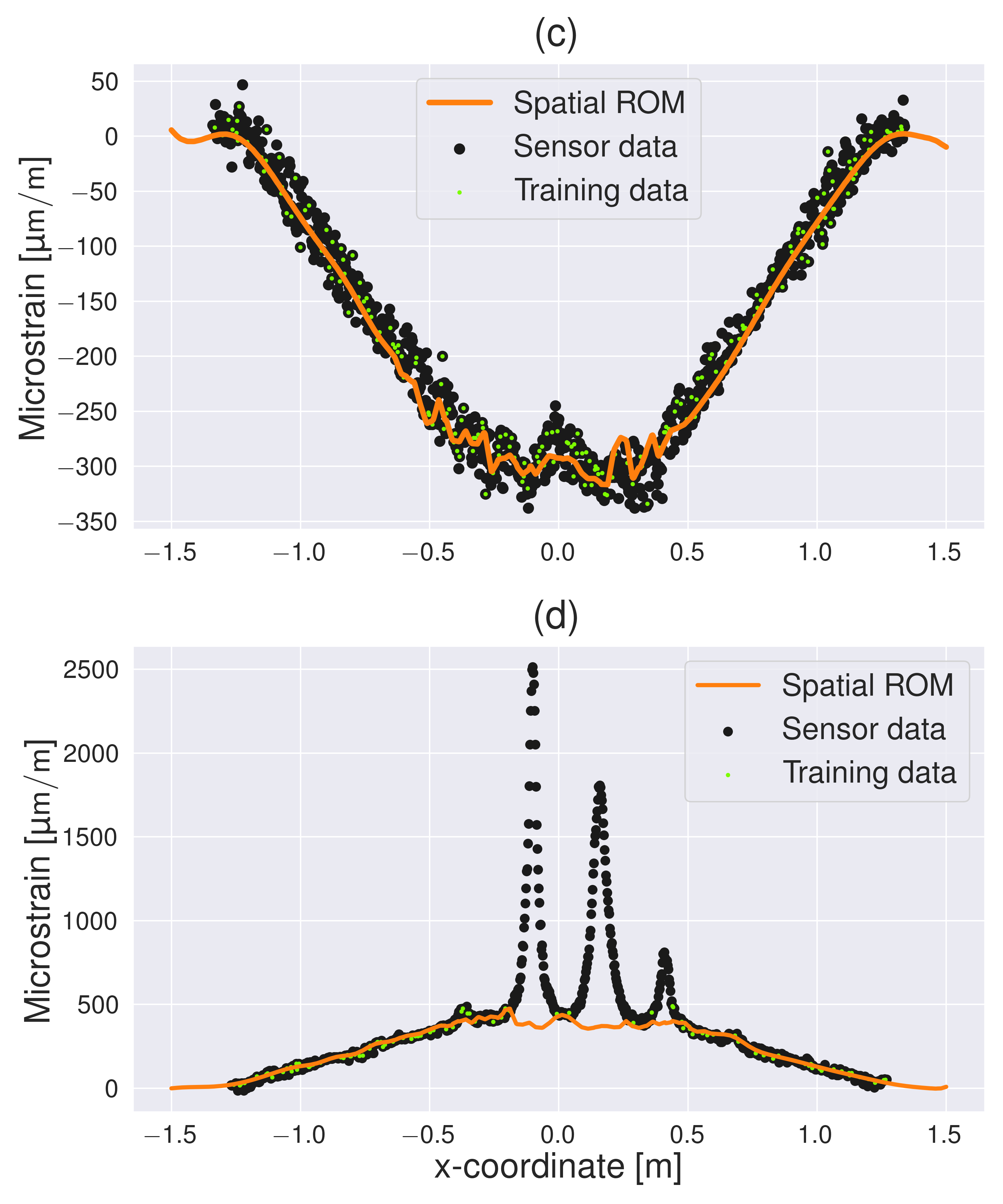}
    \includegraphics[trim={1.5cm 0 0 0},clip, width=.32\textwidth]{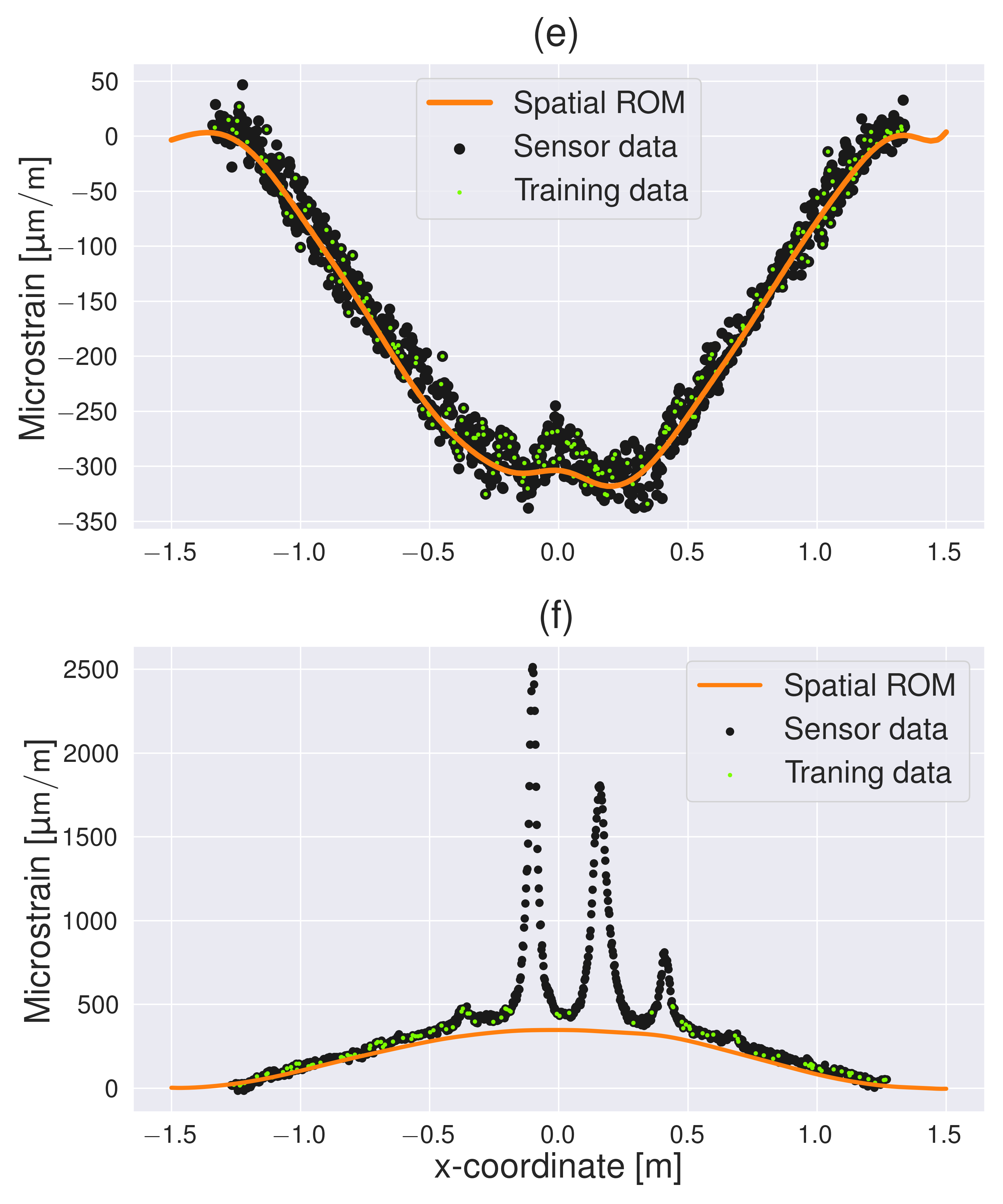}
    \caption{Strain predictions using the spatial ROM along fiber 1 and fiber 2. 
    Plots in the upper row represent the compression side (fiber 1), 
    while plots in the lower row represent the tension side (fiber 2). Different scenarios are shown as 
    (a,b) scenario 1: only compression data is involved, 
    (c,d) scenario 2: both compression and tension data are involved, 
    and (e,f) scenario 3: both compression and tension data are involved, but the weight for the experimental loss 
    is set to $w_{\mathrm{EXPs,T}}=0.01$.}
    \label{fig:ROM_1_Results}
\end{figure*}

\subsection{Results}
For this case, we consider three different scenarios. In the first scenario, the experimental loss component of 
the spatial ROM formulation includes only compression data. 
On the other hand, the experimental loss of the second scenario incorporates both compression and tension data. 
However, we exclude tension data from the damaged regions (see Fig.~\ref{figMeasurements}).  
In the last scenario, again both compression and
tension data are included. However, the loss weight of the experimental loss coming from the tension data
is reduced to $w_{\mathrm{EXPs,T}}=0.01$.

Fig.~\ref{fig:ROM_1_Results} shows the results of the three scenarios. As shown in Fig.~\ref{fig:ROM_1_Results}a, 
the spatial ROM's predictions accurately align with training and sensor data on the compression side, 
but significantly underestimate the strains on the tension side (Fig.~\ref{fig:ROM_1_Results}b). 
This disparity mainly arises from the absence of the 
tension data in the first scenario. Another reason is the structural weakening caused by cracks, 
resulting in larger deformations, which are captured by the sensors but not reflected in our simplified physical model. 
To overcome this deficiency of our physical model and obtain more accurate predictions, we additionally include sensor 
data from the tension side as depicted in Fig.~\ref{fig:ROM_1_Results}d. Now, both the tension and compression sides 
exhibit accurate alignment with the sensor data. However, the predictions are extensively biased by the noisy 
sensor data (Fig.~\ref{fig:ROM_1_Results}c and Fig.~\ref{fig:ROM_1_Results}d),  
particularly in the proximity of the beam's center, where larger strains result in higher experimental loss values.
To avoid unphysical oscillatory predictions, in the last scenario, we introduce a smaller loss weight for the tension data, 
i.e., $w_{\mathrm{EXPs,T}}=0.01$, while keeping the remaining loss weights as $1$.
By overweighing the compression data, we again obtain smoother predictions for both the compression and tension sides. 
In comparison to scenario 1, the predictions on the tension side no longer underestimate the strains, but align much better with the 
sensor data. Compared to scenario 2, the predictions are less biased by the experimental data, resulting in a smoother, non-oscillatory strain distribution.

Next, we briefly comment on the training and prediction times:
The training with the \textit{Adam} optimizer takes $\SI{25}{\second}$ on average across all scenarios. However, 
\textit{L-BFGS-B} requires $\SI{56.96}{\second}$, $\SI{171.17}{\second}$, $\SI{70.49}{\second}$
in scenario 1, scenario 2, and scenario 3, respectively. The increased training time with \textit{L-BFGS-B} 
stems from the introduction of an additional loss term involving more data points, 
as seen in scenarios 2 and 3. Moreover, the training time directly correlates with the model accuracy, 
as a more accurate model requires longer training time. 
The prediction time remains almost identical across all scenarios (approximately $\SI{0.35}{\second}$), as 
the number of trainable parameters is identical in all cases.


\begin{figure}[htbp]
    \centering
    \begin{subfigure}{0.48\textwidth}
        \includegraphics[trim={15cm 14cm 15cm 10cm},clip, width=\textwidth]{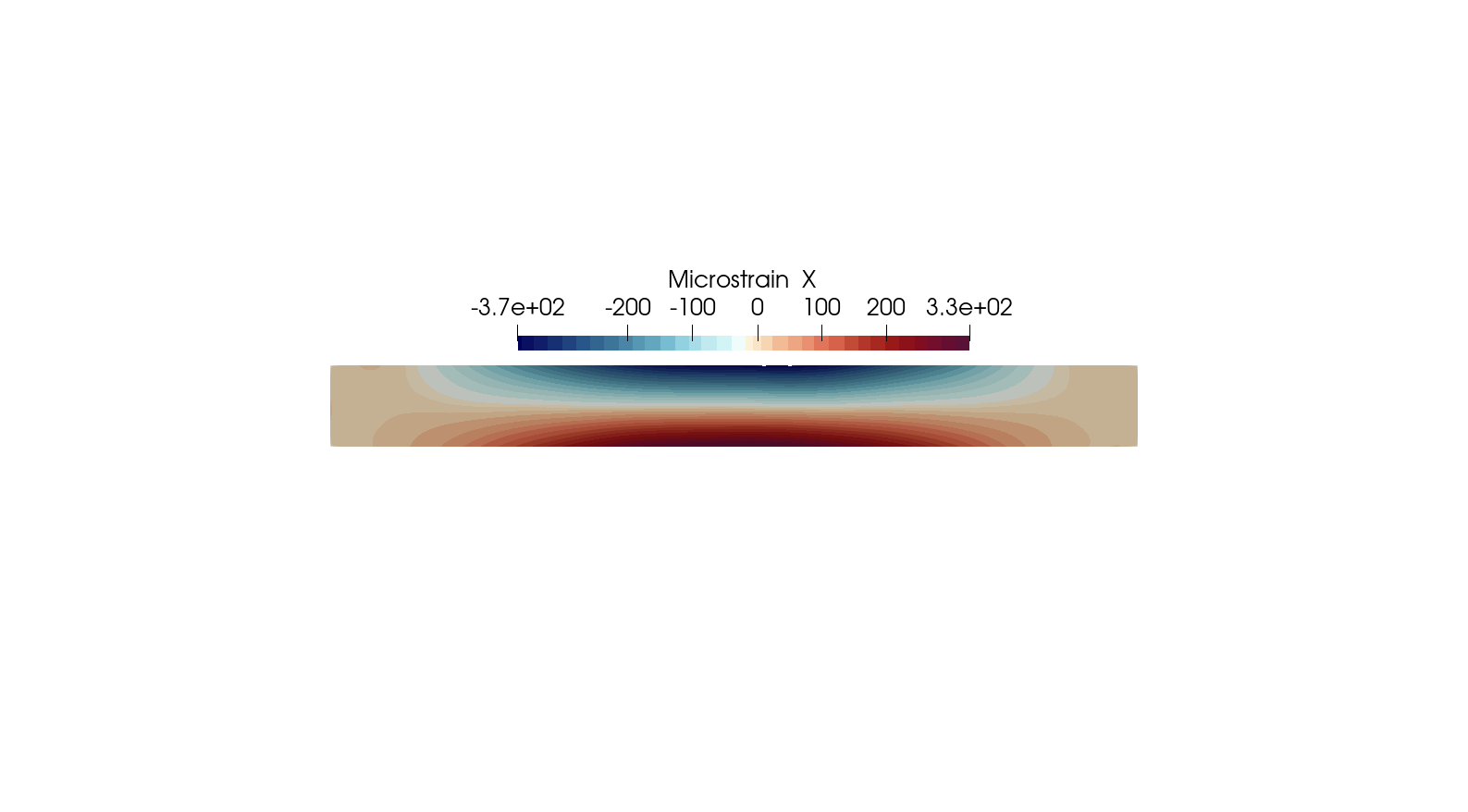}
        \caption{}
        \label{}
    \end{subfigure}%
    \vspace{0.2cm}
    \begin{subfigure}{0.48\textwidth}
        \includegraphics[trim={15cm 14cm 15cm 14cm},clip, width=\textwidth]{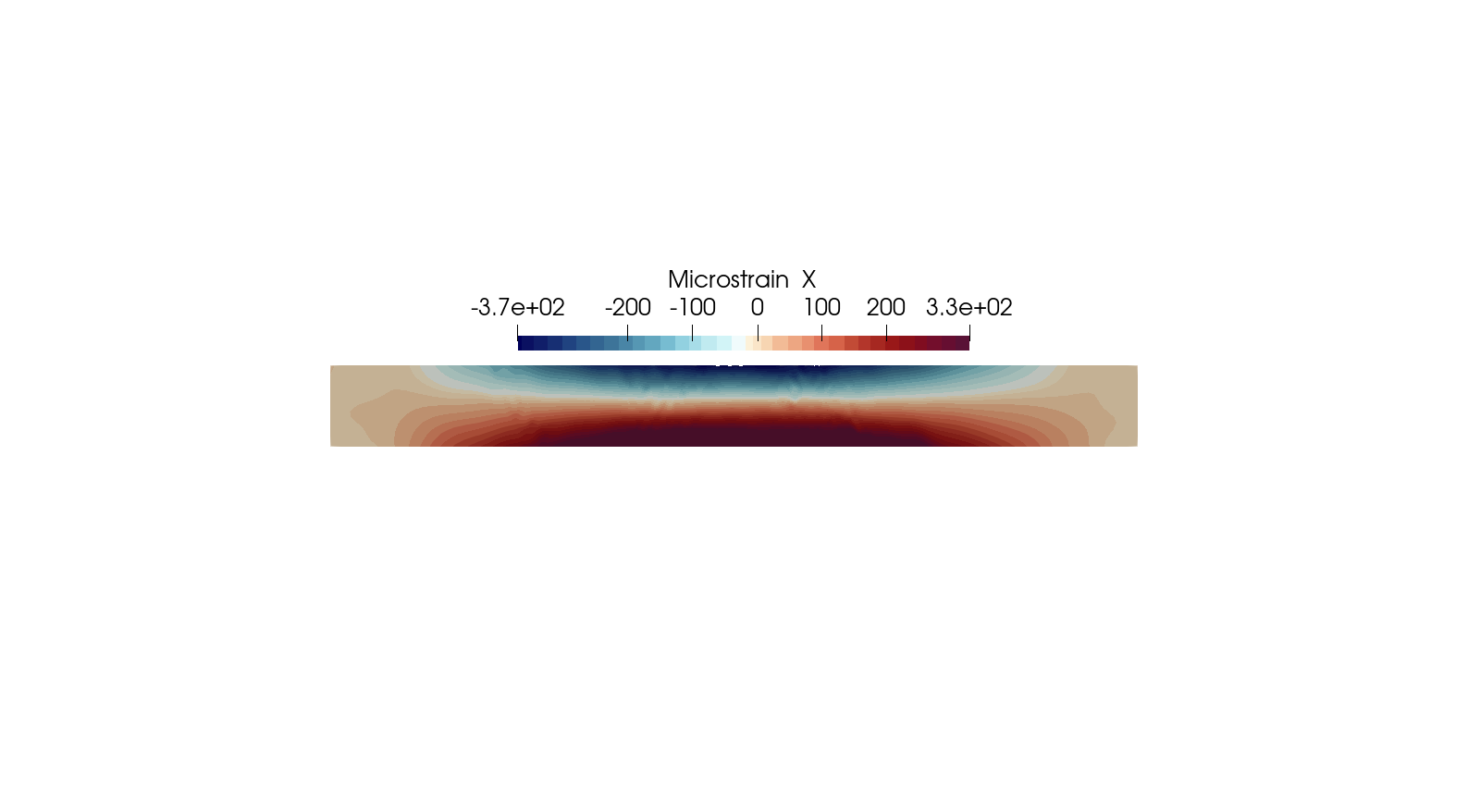}
        \caption{}
        \label{}
    \end{subfigure}%
    \vspace{0.2cm}
    \begin{subfigure}{0.48\textwidth}
        \includegraphics[trim={15cm 14cm 15cm 14cm},clip, width=\textwidth]{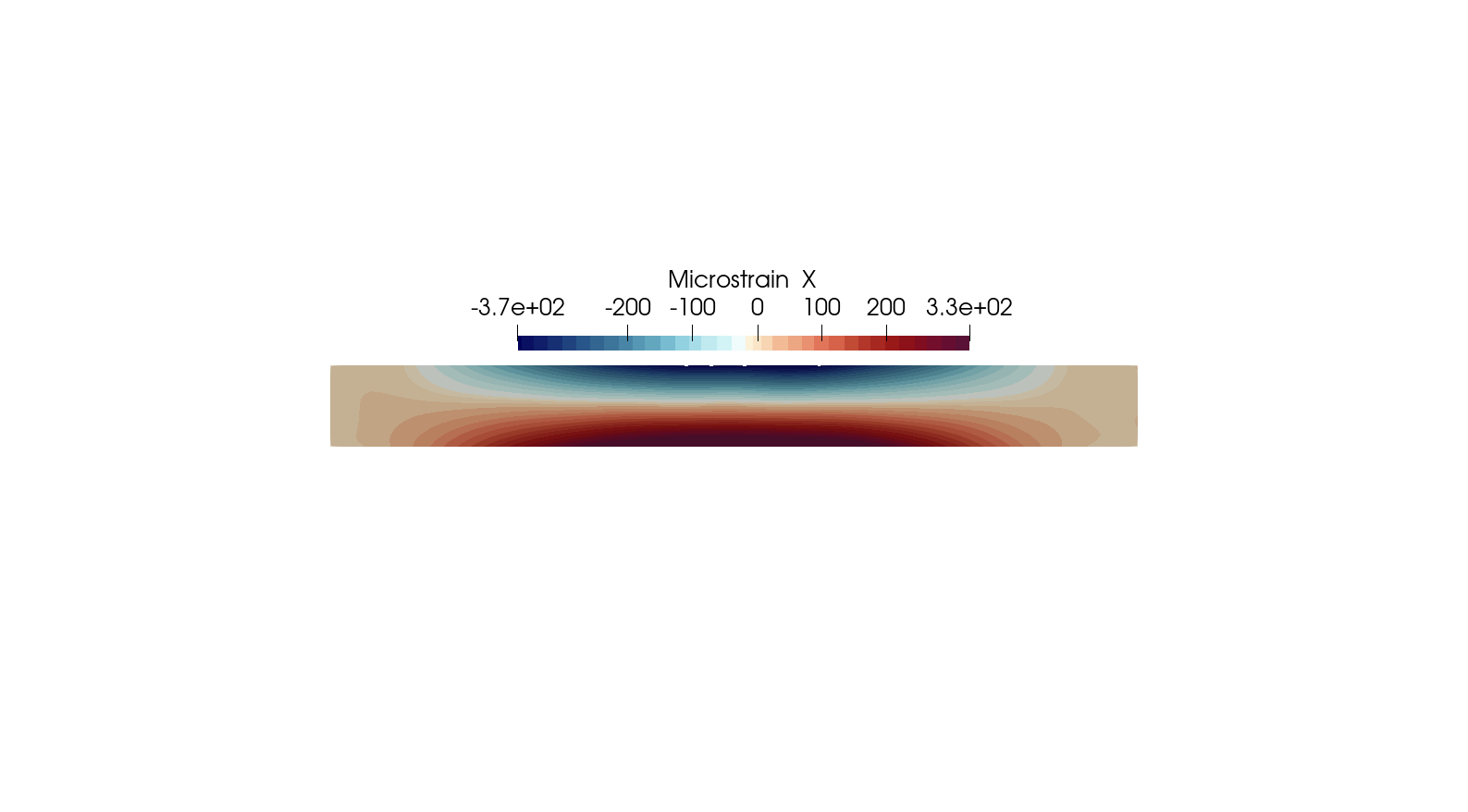}
        \caption{}
        \label{}
    \end{subfigure}
    \caption{Strain predictions using the spatial ROM within the beam domain for three scenarios: 
    (a) Scenario 1, (b) Scenario 2, (c) Scenario 3. }
    \label{fig:ROM_1_Results_all}
\end{figure}

An advantage of the PINN-based ROMs is that they enhance extrapolation capabilities and allow us to predict strains in the whole beam domain (Fig.~\ref{fig:ROM_1_Results_all}).
For the first scenario, the center of the beam exhibits nearly zero strains in the horizontal direction, 
as anticipated, and the strains are smoothly distributed in the vertical direction (Fig.~\ref{fig:ROM_1_Results_all}a). 
In the second scenario (Fig.~\ref{fig:ROM_1_Results_all}b), fluctuating strains are observed around the center line and the tension part of the beam exhibits larger strains, which can be attributed to the additional strain measurements.
In the last scenario (Fig.~\ref{fig:ROM_1_Results_all}c), the distribution of the predicted strains again closely resembles that of scenario 1, but the included data points on the tension side result in larger strains on the tension side. 
A smooth distribution is restored thanks to the weighted experimental loss.

\section{Conclusion}\label{sec:conclusion}
In this study, we presented how PINNs can be used to develop surrogate 
models for a reinforced concrete beam, which can then in turn be employed in hybrid digital twins.
The sensor data obtained from a four-point bending test, along with corresponding physics laws,
constitute the foundation for our models.

For the first surrogate model, we integrated time-dependent strain measurements acquired at fixed locations of the beam's center line
with an ODE for the motion of a harmonic oscillator. Additionally, we trained a purely data-driven model.
We observed that the method without physics-based information yields accurate predictions  
where training data is available (strong interpolation), but produces significantly poorer predictions when training data
is not available (poor extrapolation). In contrast, the physics-based surrogate model performs well in both 
interpolation and extrapolation regimes. Moreover, we identified the natural frequency of the system by exploiting the 
capability of PINNs as an inverse solver. 

In the second surrogate modeling approach, we combined the spatial experimental data with the PDE of linear elasticity 
to capture the spatial distribution of the strains throughout the beam for a fixed time. 
We examined different scenarios, where we incorporated measurement data from different locations 
into the training and varied the loss weight of the experimental data. Results show that 
involving both compression and tension data increases the accuracy of our surrogate model
specifically on the tension side, where cracks were forming during the experiment. However, the model overfits 
the data on the compression side, resulting in a fluctuating strain distribution. 
To tackle this problem, we used a smaller loss weight for the experimental data stemming from the tension side, resulting in smoother and more physical strain predictions. 
Moreover, we observe that including physics generally enhances the extrapolation capability of surrogate models 
as it allows us to make predictions within the whole beam domain, even where no sensor data is available. 

In the context of hybrid digital twins, these newly established models are promising candidates 
to be deployed as fast-to-evaluate surrogate model due to their good agreement with the measurement data.

This study unveils a clear direction for further exploration and investigation: The developed surrogate models can 
be combined into one unified model to provide strain predictions at various positions within the beam for arbitrary time 
instances. This necessitates a more sophisticated approach to combine the equations of linear elasticity, the motion of a harmonic oscillator, 
and time-dependent experimental data at different locations. 

\section{Acknowledgment}
This research paper is funded by dtec.bw – Digitalization and Technology Research Center of the Bundeswehr (project RISK.twin). 
dtec.bw is funded by the European Union – NextGenerationEU.
The authors gratefully acknowledge the computing resources provided by the Data Science
\& Computing Lab at the University of the Bundeswehr Munich.

\bibliographystyle{IEEEtran}
\bibliography{IEEEabrv,HDTPINNs}

\begin{thebibliography}{10}
\providecommand{\url}[1]{#1}
\csname url@samestyle\endcsname
\providecommand{\newblock}{\relax}
\providecommand{\bibinfo}[2]{#2}
\providecommand{\BIBentrySTDinterwordspacing}{\spaceskip=0pt\relax}
\providecommand{\BIBentryALTinterwordstretchfactor}{4}
\providecommand{\BIBentryALTinterwordspacing}{\spaceskip=\fontdimen2\font plus
\BIBentryALTinterwordstretchfactor\fontdimen3\font minus
  \fontdimen4\font\relax}
\providecommand{\BIBforeignlanguage}[2]{{%
\expandafter\ifx\csname l@#1\endcsname\relax
\typeout{** WARNING: IEEEtran.bst: No hyphenation pattern has been}%
\typeout{** loaded for the language `#1'. Using the pattern for}%
\typeout{** the default language instead.}%
\else
\language=\csname l@#1\endcsname
\fi
#2}}
\providecommand{\BIBdecl}{\relax}
\BIBdecl

\bibitem{Jiang2021}
Y.~Jiang, S.~Yin, K.~Li, H.~Luo, and O.~Kaynak, ``Industrial applications of
  digital twins,'' \emph{Philosophical Transactions of the Royal Society A:
  Mathematical, Physical and Engineering Sciences}, vol. 379, no. 2207, p.
  20200360, 2021.

\bibitem{Singh2022}
M.~Singh, R.~Srivastava, E.~Fuenmayor, V.~Kuts, Y.~Qiao, N.~Murray, and
  D.~Devine, ``Applications of digital twin across industries: A review,''
  \emph{Applied Sciences}, vol.~12, no.~11, 2022.

\bibitem{raissi2019physics}
M.~Raissi, P.~Perdikaris, and G.~Karniadakis, ``Physics-informed neural
  networks: A deep learning framework for solving forward and inverse problems
  involving nonlinear partial differential equations,'' \emph{Journal of
  Computational Physics}, vol. 378, pp. 686--707, 2019.

\bibitem{grieves2014digital}
M.~Grieves, ``Digital twin: manufacturing excellence through virtual factory
  replication,'' \emph{White paper}, vol.~1, no. 2014, pp. 1--7, 2014.

\bibitem{chinesta2020}
\BIBentryALTinterwordspacing
F.~Chinesta, E.~Cueto, E.~Abisset-Chavanne, J.~L. Duval, and F.~E. Khaldi,
  ``Virtual, digital and hybrid twins: A new paradigm in data-based engineering
  and engineered data,'' \emph{Archives of Computational Methods in
  Engineering}, vol.~27, no.~1, pp. 105--134, Jan 2020. [Online]. Available:
  \url{https://doi.org/10.1007/s11831-018-9301-4}
\BIBentrySTDinterwordspacing

\bibitem{torzoni2024}
M.~Torzoni, M.~Tezzele, S.~Mariani, A.~Manzoni, and K.~E. Willcox, ``A digital
  twin framework for civil engineering structures,'' \emph{Computer Methods in
  Applied Mechanics and Engineering}, vol. 418, p. 116584, 2024.

\bibitem{von2023hybrid}
M.~{von Danwitz}, T.~T. Kochmann, T.~Sahin, J.~Wimmer, T.~Braml, and A.~Popp,
  ``Hybrid digital twins: A proof of concept for reinforced concrete beams,''
  \emph{Proceedings in Applied Mathematics and Mechanics}, vol.~22, no.~1, p.
  e202200146, 2023.

\bibitem{braml2022b}
T.~Braml, J.~Wimmer, and Y.~Varabei, ``{E}rfordernisse an die {D}atenaufnahme
  und -verarbeitung zur {E}rzeugung von intelligenten {D}igitalen {Z}willingen
  im {I}ngenieurbau,'' in \emph{Innsbrucker Bautage 2022 : Festschrift zum 60.
  Geburtstag von Univ.-Prof. Dr.-Ing. Jürgen Feix}, ser. Innsbruck - Massivbau
  und Brückenbau, J.~Berger, Ed., vol.~7.\hskip 1em plus 0.5em minus
  0.4em\relax Innsbruck: Studia Verlag, 2022, pp. 31--49.

\bibitem{lawrence1993}
J.~Lawrence, \emph{Introduction to Neural Networks}.\hskip 1em plus 0.5em minus
  0.4em\relax USA: California Scientific Software, 1993.

\bibitem{sahin2024}
T.~Sahin, M.~Von~Danwitz, and A.~Popp, ``Solving forward and inverse problems
  of contact mechanics using physics-informed neural networks,'' \emph{Advanced
  Modeling and Simulation in Engineering Sciences}, vol.~11, no.~1, p.~11,
  2024.

\bibitem{lu2021deepxde}
L.~Lu, X.~Meng, Z.~Mao, and G.~E. Karniadakis, ``{DeepXDE}: A deep learning
  library for solving differential equations,'' \emph{SIAM Review}, vol.~63,
  no.~1, pp. 208--228, 2021.

\end{thebibliography}


\end{document}